\documentclass[12pt,preprint]{aastex}

\usepackage{lscape}

\slugcomment{Accepted by \emph{ApJ Letters}}

\shorttitle{M31 RCBs}
\shortauthors{Tang et al.}

\begin{document}


\title{R Coronae Borealis Stars in M31 from the Palomar Transient Factory}


\author{Sumin Tang\altaffilmark{1,2}, Yi Cao\altaffilmark{2}, Lars Bildsten\altaffilmark{1}, Peter Nugent\altaffilmark{3,4}, 
Eric Bellm\altaffilmark{2}, Shrinivas R. Kulkarni\altaffilmark{2}, Russ Laher\altaffilmark{5}, David Levitan\altaffilmark{2}, Frank Masci\altaffilmark{6}, Eran O. Ofek\altaffilmark{7}, Thomas A. Prince\altaffilmark{2}, Branimir Sesar\altaffilmark{2} \& Jason Surace\altaffilmark{5} }

\altaffiltext{1}{Kavli Institute for Theoretical Physics, University of California, Santa Barbara, CA 93106, USA}
\altaffiltext{2}{Division of Physics, Mathematics, \& Astronomy, California Institute of Technology, Pasadena, CA 91125, USA}
\altaffiltext{3}{Computational Cosmology Center, Lawrence Berkeley National Laboratory, 1 Cyclotron Rd., Berkeley CA 94720, USA}
\altaffiltext{4}{Department of Astronomy, University of California, Berkeley, California 94720-3411, USA}
\altaffiltext{5}{Spitzer Science Center, California Institute of Technology, Pasadena, CA 91125, USA}
\altaffiltext{6}{Infrared Processing and Analysis Center, California Institute of Technology, Pasadena, CA 91125, USA}
\altaffiltext{7}{Benoziyo Center for Astrophysics and the Helen Kimmel center for planetary science, Weizmann Institute of Science, 76100 Rehovot, Israel}


\begin{abstract}
We report the discovery of R Coronae Borealis (RCB) stars in the Andromeda galaxy (M31) using the Palomar Transient Factory (PTF).
RCB stars are rare hydrogen-deficient, carbon-rich supergiant variables, most likely the merger products of two white dwarfs.
These new RCBs, including two confirmed ones and two candidates, are the first to be found beyond the Milky Way and the Magellanic Clouds.
All of M31 RCBs showed $>1.5$ mag irregular declines over timescales of weeks to months.
Due to the limiting magnitude of our data ($R\approx21-22$ mag),
these RCB stars have $R\approx19.5$ to 20.5\,mag at maximum light, corresponding to $M_R= -4$ to $-5$, 
making them some of the most luminous RCBs known.
Spectra of two objects show that they are warm RCBs, similar to the Milky Way RCBs RY Sgr and V854 Cen.
We consider these results, derived from a pilot study of M31 variables,
as an important proof-of-concept for the study of rare bright variables in nearby galaxies with the PTF or other synoptic surveys.
\end{abstract}





\section{Introduction}
R Coronae Borealis (RCB) stars are carbon-rich, hydrogen-poor supergiants with large irregular declines up to
$8-9$\,mag in their optical light curves due to dust formation episodes (see e.g., reviews by Clayton 1996, 2012). 
Spectroscopic analyses yield atmospheric abundances that are predominantly helium with $\approx1$\% carbon for the majority of RCBs (Asplund et al. 2000).
Two evolutionary scenarios have been suggested: mergers of two white dwarfs (WDs), or a final helium shell flash in the
central star of a planetary nebula (Webbink 1984; Iben et al. 1996). 
Recent observations, especially the low $^{16}$O/$^{18}$O ratio ($\sim1$) observed in some RCBs, 
favor the WD-WD merger model (Clayton et al. 2005, 2007; Garc{\'{\i}}a-Hern{\'a}ndez et al. 2010; Staff et al. 2012; Menon et al. 2012).
Hence, RCBs are possibly low-mass analogs to the double-degenerate (DD) model of Type Ia supernovae (SNe; Iben \& Tutukov 1984). 
Study of these systems thus provides information on the rate of WD binary mergers, 
potentially shedding light on the SNe Ia rate via the DD channel.

Despite the fact that RCBs are among the brightest stars ($M_V \sim -3$ to $-5$ in quiescence; Tisserand et al. 2009), 
and have large amplitude variations, their population is still poorly characterized.
There are $\sim$5000 RCBs expected in the Milky Way extrapolated from the LMC RCB population (Alcock et al. 2001), 
but only 76 have been found, and their distribution is heavily biased by the microlensing survey field positions (Clayton 2012; Tisserand et al. 2012). 
Miller et al. (2012) and Tisserand et al. (2012) found over 20 new galactic RCBs in the ASAS dataset (Pojma{\'n}ski 1997), but more work remains.
Two major challenges --- the heavy extinction and crowding along the galactic plane --- make it fairly difficult
to obtain a full census of RCBs (or any other variable sources) in our own galaxy.

As the closest large spiral galaxy similar to the Milky Way, M31 provides an ideal laboratory for the census of RCBs in a spiral galaxy. 
The Palomar Transient Factory (PTF; Law et al. 2009) has been monitoring M31 intensively since 2009.
This dataset offers a unique resource for studying bright variables in M31 on timescales from days to years. 
We present here our first results of a RCB search in M31.  
We describe the PTF survey data in \S 2.
Our RCB candidate selection procedure and RCB light curves are presented in \S 3.
Follow-up spectra and analyses are presented in \S 4.
Our main results are summarized in \S 5.

\section{PTF M31 Data}
The PTF (Law et al. 2009; Rau et al. 2009) uses the Samuel Oschin 1.2-m Schmidt telescope at the Palomar Observatory. 
It uses the PTF CFH12k camera with 11 CCDs, each 2K$\times$4K pixels (Rahmer et al. 2008).
The pixel size is  1.01\arcsec/pixel and the total field of view is 7.2\,deg$^{2}$.
PTF has been monitoring M31 regularly  since Aug 2009 with a typical cadence of twice per night when M31 is accessible.
In several nights, M31 has been observed as much as 150 times a night.
Over 80\% images were taken in Mould$-$R filter (the transmission curve is available from Ofek et al. 2012b), 
and the remaining were in SDSS $g'$ and $\rm{H\alpha}$ filters.
All broad-band images have an integration time of 60\,s.
In this paper, we used R-band data from Aug 2009 to Oct 2012,  
which consists of 1394 images, 
including 1189 images in PTF field 100043 and 205 images in field 4445.
Figure 1 shows a PTF mosaic image of the M31 field.

All images were processed using the PTF photometric pipeline hosted by the Infrared Processing and Analysis Center (IPAC; Grillmair et al. 2010).
Images are processed using standard reduction procedures, including de-biasing, flat-fielding, and astrometric calibration (Laher et al. in prep). 
The absolute photometric calibration is accurate to $\approx$2\% (Ofek et al. 2012a).
Source matching and relative photometric calibration generates the final light curves and refines the calibration to $<$1\% level (Levitan et al. 2011, Levitan et al. in prep).
The typical image Full Width at Half Maximum is $\approx2$\arcsec.
The typical photometric uncertainty at the bright end ($R<16$) is $\sigma<0.01$\,mag,
which increases to $\sigma \approx0.2$\,mag at $R\approx20-21$ (the 5-$\sigma$ limiting magnitude).
M31 has a distance modulus of 24.4 (Vilardell et al. 2010), 
so we probe the photometric evolution of stars brighter than $-4$.

\section{RCB Candidate Selection}
We used two approaches to select initial candidates.
The first approach makes use of optical colors from SDSS DR9 (Ahn et al. 2012) 
and chi-square per degree of freedom of the light curves.
Most RCBs have effective temperatures of $T_{\rm eff}=5000-7000$K,
while the major contaminants in this search, i.e., non-RCBs with irregular dimming-like light curves, 
are cool red stars (including irregular red giants, DY Per or other carbon stars), or blue quasars.
The SDSS catalog does not cover the bulge and disk regions of M31 due to crowding,
and therefore this approach only works in the outskirt halo region, as shown in Figure 1.

There are 160,071 objects with at least 20 measurements in the light curves of our chosen search field with SDSS colors available.
For each light curve, we compute the chi-square per degree of freedom as 
$\chi^2 = \frac{1}{n-1}\sum_{i=1}^{n}\frac{(R_i-\langle R \rangle)^2}{\sigma_i^2}$,
where $n$ is the number of detections, $R_i$ is the R-band magnitude, $\langle R\rangle$ is the mean magnitude in R-band, and $\sigma_i$ is the photometric error.
There are 10,198 light curves with $\chi^2>5$.
We removed most blue and red contaminants (i.e., quasars and red giants) by requiring $u-g>1$, $0<g-r<1$, and $ r>18$ (i.e., $M_r>-6.4$).
As listed in Table 1, SDSS colors were obtained before PTF images and it is unknown whether the candidates were at maximum brightness at the time of the SDSS epochs. 
RCBs during decline phases could be redder due to the dust, and thus RCBs at decline phases at SDSS epochs might be missed by our color cuts.
The number of variable sources reduced to 1,547 after these color and magnitude cuts.
We further applied a constraint on the amplitude of variation of $\Delta R = R_{\rm min} - R_{\rm median}>1$,
where $R_{\rm min}$ is the faintest magnitude, and $R_{\rm median}$ is the median magnitude in the light curve.
A total of 141 objects were left as initial candidates.

To cover the central crowded region of M31 without SDSS colors, we took a second approach based on light curves alone.
To avoid excess dubious variations due to blending, only those light curves with less than 20\% of blended detections are included.
This process loses objects in crowded regions, and therefore the search is still biased towards selecting variables in the outskirt region.
A decline episode of a RCB star usually lasts weeks to months (Clayton 2012).
With the cadence of twice per night, we are likely to see consecutive points in any decline episodes of RCBs, which could be used as a feature to distinguish RCBs and other variable stars.

For each light curve, we defined a measurement of consecutive dip points, $n_{dip, 4\sigma}$, as follows.
A detection is defined as in a `dip' state if it is 4-$\sigma$ fainter than the median magnitude of the light curve, where $\sigma$ is the photometric uncertainty of the detection.
Non-detections are not included in the calculation. 
We defined $n_{dip, 4\sigma}$ as the number of consecutive 4$\sigma$ dip detections in the light curve.
Here are some examples: If a source has two measurements in `dip' state and the two are not consecutive in the light curve, then $n_{dip, 4\sigma}=0$;
If a source has two measurements in `dip' state and the two are consecutive in the light curve, then $n_{dip, 4\sigma}=1$;
If a source has three measurements in `dip' state and only two out of the three are consecutive in the light curve, then $n_{dip, 4\sigma}=1$.
We then divided all sources into 20 magnitude bins, each containing the same number of sources,
and compared the distribution of $n_{dip, 4\sigma}$ in each magnitude bin.
RCBs are expected to have consecutive detections in the `dip' state, and thus have larger $n_{dip, 4\sigma}$ values.
We selected the 6-$\sigma$ outliers of the $n_{dip, 4\sigma}$ distribution in each magnitude bin,
where $\sigma$ is the standard deviation of the distribution,
and identified 387 sources as initial candidates in this second approach.

For the initial candidates derived from the above two approaches, we extracted their coadded light curves as follows.
We selected images with seeing better than 3$^{\prime\prime}$ and stacked every $\approx30$ epochs with SWarp (Bertin et al. 2002)
to obtain a time series of co-added images that reach $22-22.5$\,mag. 
Difference images were calculated using HOTPANTS\footnote{HOTPANTS can be found at http://www.astro.washington.edu/users/becker/hotpants.html} 
using the first co-added image as a reference. 
We measured the flux of each target on each subtraction image with an aperture of twice the seeing radius. 
Magnitudes were then converted from these counts plus counts of their counterparts on the reference image with the same aperture.

After a visual inspection of their light curves and images (both single epoch and coadded ones),
we found only six candidates which showed convincing $>1.5$ mag irregular declines and thus considered likely RCBs for spectroscopic follow-up.
Their positions are marked in Figure 1, and their properties are listed in Table 1.
The first three, PTF-M31-RCB-1 (hereafter RCB-1, and so on), RCB-2 and RCB-Cand-1,
are from the 141 candidates of the first approach using SDSS colors.
The other three, as well as RCB-Cand-1, are from the 387 candidates of the second approach.
The declines of RCB-1 and RCB-2 are deep with mostly non-detections in the light curves,
and both objects were missed  in the second approach due to that non-detections are not included in our calculation of $n_{dip, 4\sigma}$.
Finding charts of the first four from Table 1 are shown in Figure 1,
and their light curves are shown in Figure 2.
The 521 objects we rejected were mostly pulsating stars, AGNs, or observational defects such as blending.

\section{Spectroscopy}

\subsection{Observations}
The spectroscopic observations were obtained between Sep and Dec 2012 using the Double Spectrograph (DBSP; Oke \& Gunn 1982) mounted on the Palomar 5.1-m Hale telescope (P200), 
and the Deep Imaging Multi-Object Spectrograph (DEIMOS; Faber et al. 2003) mounted on the Keck II 10-m telescope.
RCB-2, PTFJ0041+4038 and PTFJ0046+4042
were observed with DBSP using gratings 600/4000\AA\ in the blue and 158/7500\AA\ in the red with 20\,min exposure time;
RCB-Cand-1 was observed with DBSP using gratings 600/4000\AA\ in the blue and 316/7500\AA\ in the red with 25\,min exposure time.
RCB-1, RCB-2 and RCB-Cand-1 
were observed with DEIMOS with  600ZD grating (600/7500\AA); 
The exposure time was 40\,min for RCB-1, and 20\,min for RCB-2 and RCB-Cand-1.

All data were reduced with standard IRAF routines. 
RCB-Cand-2 was too faint ($R>21.5$ mag) during the follow-up observation, and therefore we did not obtain a spectrum.

\subsection{Line-of-Sight Velocities}
M31 has a large negative systemic velocity of $\approx-300$ km/s and a high rotation rate of up to $270$ km/s  in the disk (Rubin \& Ford 1970; Sofue \& Kato 1981),
while most foreground Milky Way stars have radial velocities of $>-150$ km/s (Hartmann \& Burton 1997).
Thus it is relatively straightforward to verify the M31 membership of the RCB candidates via radial velocity measurements.

We used the following lines to measure heliocentric radial velocities when available:    
Ca\,{\sc ii} triplet (8498.03 \AA, 8542.09 \AA \ and 8662.14 \AA), 
Balmer lines (H$\rm \alpha$, H$\rm \beta$, H$\rm \gamma$, H$\rm \delta$ and H$\rm \epsilon$), 
Si\,{\sc ii} (6347.09 \AA \ and 6371.36 \AA),
and Ca\,{\sc ii} K \& H (3933.66 \AA, 3968.47 \AA).
The radial velocities were measured by fitting a Gaussian to the lines.
For objects observed with both DBSP and DEIMOS, we used the DEIMOS spectra for the blue part ($<5000$\AA) which have better S/N ratio and higher resolution.
All five candidates with spectra have multiple lines in the spectra available for radial velocity fitting.
For each object, we then defined the radial velocity, $V_{LOS}$, as the weighted mean of radial velocities derived from different lines,
while its uncertainty is given by the standard deviation of radial velocities derived from different lines.
As listed in Table 1, all the five candidates have radial velocities $<-200$ km/s, and are consistent with being members of M31.

\subsection{Spectroscopic Confirmation}
The two major characteristic features of RCB star spectra are hydrogen deficiency and carbon richness.
Cooler RCBs ($T_{\rm eff}<6000$ K) show strong carbon bands,
while warmer RCBs  ($T_{\rm eff}>6000$ K), such as RY Sgr, can appear almost featureless (Clayton 2012),
which makes it more challenging to identify as in our case.
In addition, most RCBs have a low abundance of $^{13}\rm C$, unlike other carbon stars.
A rich set of optical spectra for over 40 RCBs can be found in Tisserand et al. (2012).

 We have found that two of the spectroscopically observed stars have spectra similar to known warm RCBs, 
 and consider them as confirmed RCB stars in M31.
 Their light curves are shown in Figure 2, and their spectra are shown in Figures 3 and 4. 
 Spectra of both objects have no carbon bands, and therefore the relative strengths of $^{12}\rm C$ and $^{13}\rm C$ are unknown.
 RCB-1 has no H$\alpha$, and its spectrum is very similar to a known warm RCB in the Milky Way, RY Sgr.
 The spectrum of RCB-2 is very similar to another warm RCB in the Milky Way, V854 Cen, with weak H$\alpha$.
 This is not surprising, as we are selecting the brightest RCBs ($M_R\sim -4$ to $-5$), 
 which are likely to be warmer, as observed in the Magellanic Clouds (Alcock et al. 2001; Tisserand et al. 2009).

 The nature of the other two RCB candidates is not entirely clear.
 Their light curves are shown in Figure 2.
 RCB-Cand-1 showed a $>1.9$ mag dip which lasted for two months; it is otherwise at a constant maximum light level.
 As shown in Figures 3 and 4, it has an almost featureless spectrum with some hydrogen.
Tisserand et al. (2012) found that the main families of objects with RCB-like light curves but later rejected spectroscopically
 are Mira variables, hydrogen emission line stars, or carbon stars with high abundance of $^{13}$C.
 The spectrum of RCB-Cand-1 does not look like any of the above.
 Besides, it has a radial velocity of $\approx-220$km/s, and thus is probably a member of M31 with absolute magnitude of $M_R=-4.6$, consistent with being a warm RCB.
However, it is hard to confirm it with the current spectrum which shows no carbon bands, and does not look like any known warm RCBs. 
 RCB-Cand-2 showed a 1.6 mag dip with relatively faster decline and slower recovery, which is typical for RCB stars.
 A spectrum is still needed for identification.

The other two candidates are rejected due to clear evidence of hydrogen or the existence of $^{13}\rm C$, as summarized in Table 1.
Both candidates have radial velocities $<-300$ km/s, and are most likely M31 members.
PTFJ0041+4038 is the brightest one among our candidates with a maximum magnitude of $R=18.1$, which corresponds to $M_R=-6.3$ at the distance of M31.
It shows C2 bands, $^{13}\rm C$ and hydrogen, and could be a carbon-rich AGB star.
PTFJ0046+4042 shows strong Balmer emission lines, and might be a Be star, 
although the $>2$ mag decline is unusual for a Be star (Mennickent et al. 2002).

\section{Summary and Discussion}
We have discovered two new RCB stars and two RCB candidates in M31 using PTF data.
These RCBs are the first discovered beyond the Milky Way and Magellanic Clouds.
Follow-up spectra show that these are warm RCBs,
similar to RY Sgr and V854 Cen, 
consistent with our selection for the most luminous RCBs,
which are expected to be warmer (Tisserand et al. 2009).

The population and distribution of RCBs in the Milky Way is still poorly understood.
As the nearest spiral galaxy similar to the Milky Way, M31 provides a unique opportunity.
Our pilot search is mostly sensitive to the outskirt region, and only 1 out of 6 candidates with RCB-like light curves are located in M31's disk (which turned out to be a non-RCB after follow-up spectroscopy).
There are two issues limiting our search.
First, we used SDSS colors in the search to eliminate red giant variables and quasars.
These colors are not available for the bulge and disk region of M31.
Second, we limited our search to sources which are mostly not blended, and as a result, sources in crowded environments are not included.
Both are related to the challenge of photometry in a crowded field.

A future study would benefit from including the full PTF M31 field with improved difference imaging and PSF photometry (Masci et al. in prep),
and making use of other available catalogs for optical colors (e.g., Massey et al. 2006).
As demonstrated by Tisserand (2012) and Miller et al. (2012),
IR colors and machine-learning algorithms can greatly improve the efficiency of searches for RCBs,
and will likely increase the number of newly discovered RCBs in M31.
In LMC, 5 out of 19 known RCBs have V$\le$14.3\,mag (Clayton 2012), 
which corresponds to V$\le20.2$\,mag and R$\lesssim$20\,mag for F to G type stars at the distance of M31.
Similar to that in the Milky Way, the total number of RCBs in M31 is expected to be about a few thousand,
and if the luminosity distribution is similar to that in LMC, over one thousand RCBs are probably brighter than $R=20$\,mag.
A certain fraction of them are probably enshrouded by dust or too blended to obtain good light curves, and some others may not show decline episodes during PTF observations.
Assuming a conservative detection efficiency of 10\% of bright RCBs (R$\le$20\,mag), 
we expect to find a few dozen of them in the PTF in future searches.
This will truly improve our understanding of the RCB population in a spiral galaxy.

\acknowledgments
We thank Geoff Clayton for pointing out the similarities between PTF-M31-RCB-1 (2) and RY Sgr (V854 Cen), 
providing spectra of RY Sgr and V854 Cen, and many helpful discussion.
We thank the anonymous referee for comments that have helped to improve this paper.
We thank Sagi Ben-Ami and Dong Xu for reducing two candidate spectra,
and Adam Miller for helpful discussion.
This work was supported by the National Science Foundation under grants PHY
11-25915 and AST 11-09174.
E.O.O. is incumbent of the Arye Dissentshik career development chair and
is grateful to support by a grant from the Israeli Ministry of Science.



{\it Facilities:} \facility{PO:1.2m (PTF)}, \facility{Hale(DBSP)}, \facility{Keck (DEIMOS)}.

\clearpage

\begin{landscape}
\begin{table} \tiny
\begin{center}
\caption{List of newly confirmed, candidate, and rejected RCB stars in M31 from PTF.\label{tbl-2}}
\begin{tabular}{lccccccccccc} 
\tableline\tableline
 & & & PTF & & &&& SDSS &&& \\
ID & IAU Name & $R_{max}  $ & $\Delta$R &  $V_{LOS} $  &  u & g & r & i & z  & Epoch & Remarks  \\
  &   & (mag)  & (mag)  &  (km s$^{-1}$) &   &  &  &  &  & MJD &  \\
\tableline
New RCB stars: \\
PTF-M31-RCB-1 & PTF1J004755.66+400732.6 & 20.0 & $>1.9$  & $-321\pm20$  & 23.53 &  20.80 & 20.19 & 20.20 & 20.27 & 54010 & Spectrum similar to RY Sgr \\
PTF-M31-RCB-2 & PTF1J004915.08+413844.5 & 19.4 & $>2.5$ & $-278\pm44$ & 21.63 & 20.46 & 19.89 & 19.77	 & 19.69 & 52553 & Spectrum similar to V854 Cen \\
\tableline
New RCB star candidates: \\
PTF-M31-RCB-Cand-1 & PTF1J004521.00+405213.8 & 19.8  & $>1.9$  & $-221\pm84$  & 22.37 & 20.51 & 19.95 & 19.60 & 19.40 & 52553 & Featureless, some H\\
PTF-M31-RCB-Cand-2 & PTF1J004511.93+410516.5 & 20.3 & 1.6 &  N/A  & 25.47 & 22.88 & 21.26 & 20.81 & 20.29 & 52553 & No spectrum available \\
\tableline
Rejected RCB candidates: \\
PTFJ0041+4038 & PTF1J004123.20+403842.5 & 18.1 & $3$  & $-475\pm20$ & N/A & N/A & N/A & N/A & N/A& N/A&  $^{13}\rm C$, H, too bright \\ 
PTFJ0046+4042 & PTF1J004620.75+404209.0 & 20.1 &  $>2$ & $-349\pm37$  & 21.01 & 20.68 & 20.07 & 19.82 & 19.61 & 52553 & Balmer emission lines \\ 
\tableline
\end{tabular}
\end{center}
\end{table}
\end{landscape}



\begin{figure}
\epsscale{1.0}
\plotone{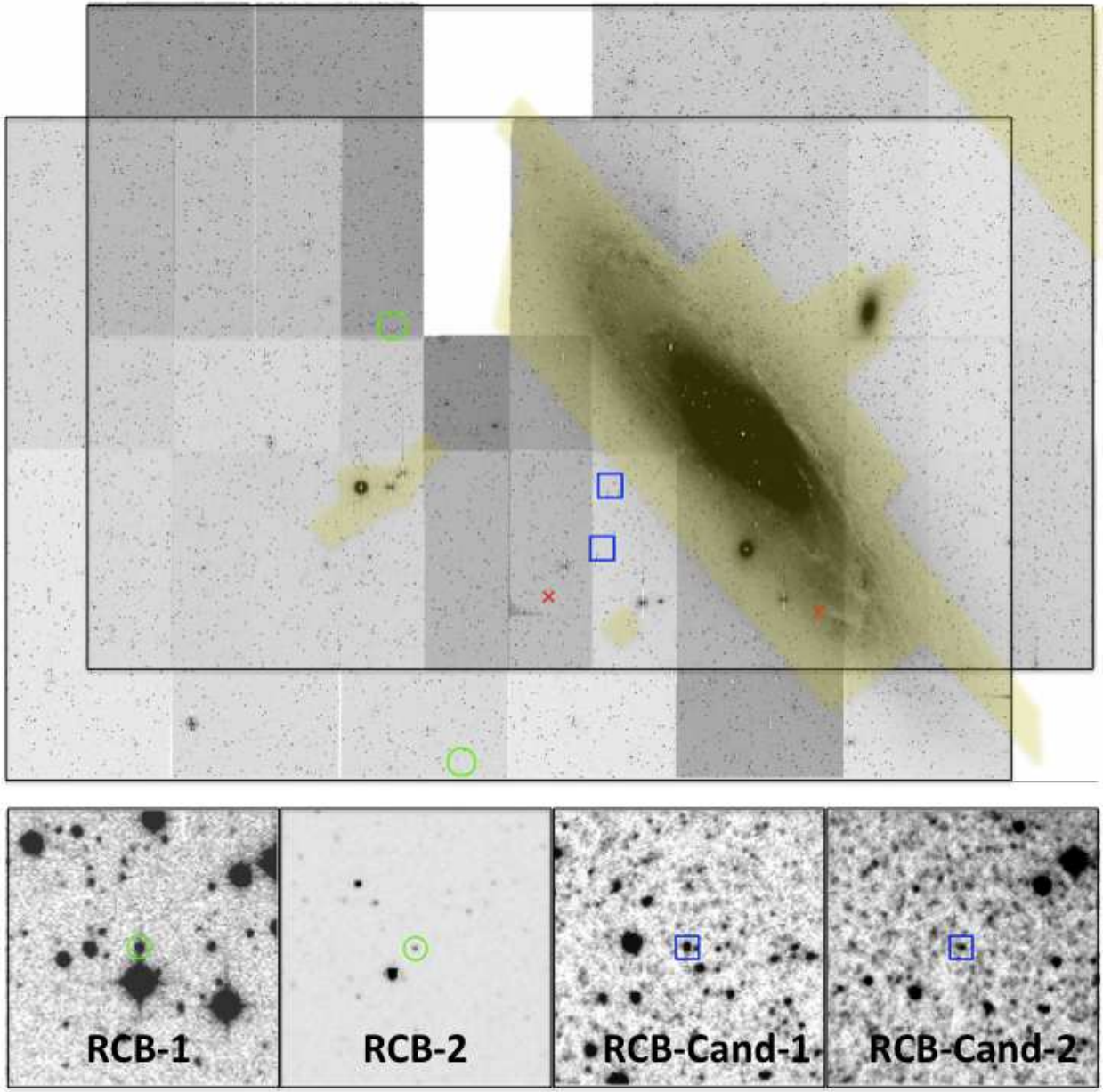}
\caption{PTF image of the M31 field and finding charts of the newly discovered RCBs.  
North is up and East is to the left. 
Upper: mosaic image consisting of two overlapping M31 fields, i.e., field 100043 (lower-left) and field 4445 (upper-right).
The upper-middle blank region is due to one broken CCD chip (CCD 3).
Regions where the SDSS catalog is not available are marked in yellow,
including the M31 disk and M110, the upper-right corner where no SDSS image is available, 
and two smaller regions in the south-east of M31. 
The green circles denote positions of the confirmed RCBs,
blue boxes are RCB candidates, and red crosses are the ones rejected as RCBs by follow-up spectra.
Lower: finding charts of four RCBs or RCB candidates in PTF R-band. Each finder is 2\arcmin$\times$2\arcmin.
\label{fig1}}
\end{figure}

\begin{figure}
\epsscale{1.0}
\plotone{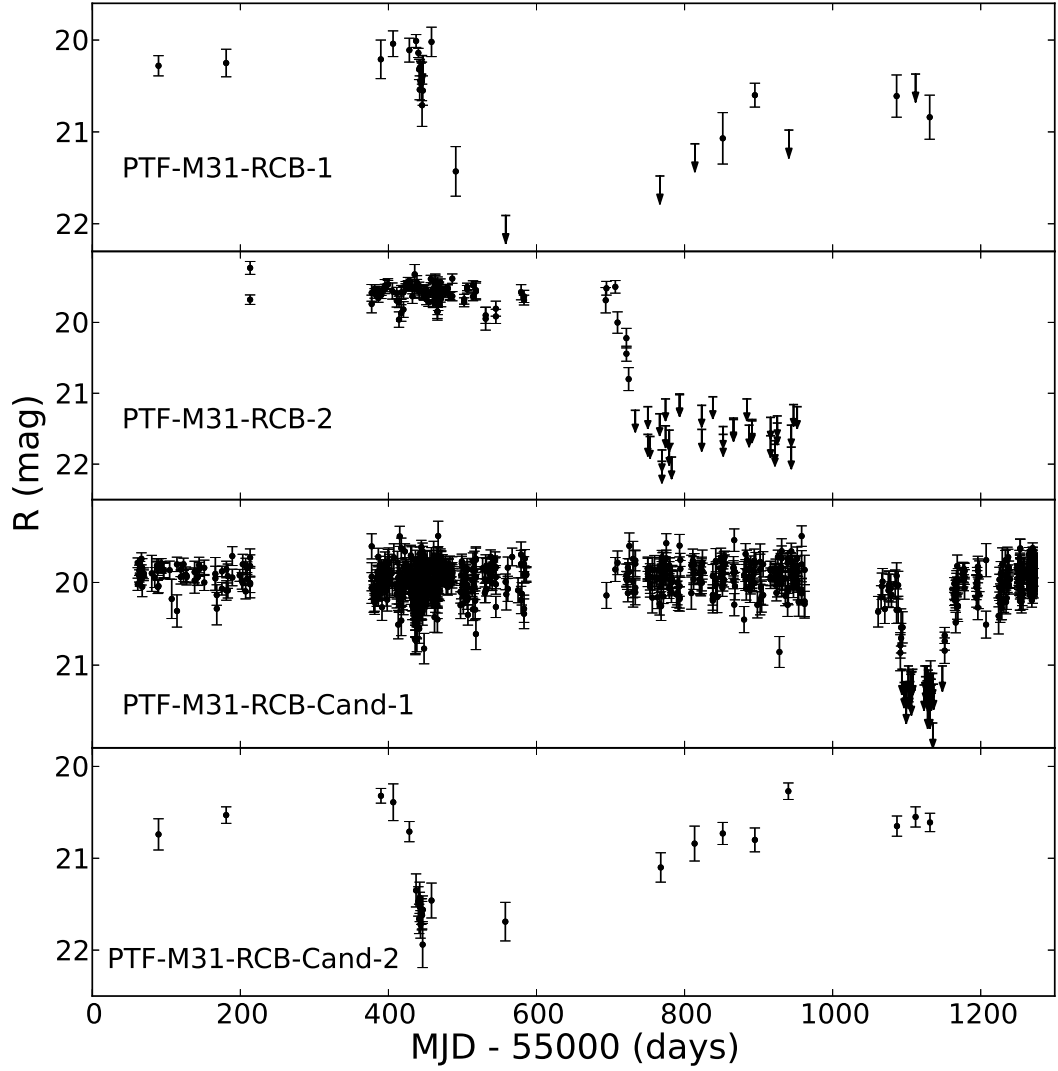}
\caption{PTF R-band light curves of the newly confirmed and candidate RCBs in M31. Downward arrows indicate upper limits. \label{fig2}}
\end{figure}

\clearpage


\begin{figure}
\plotone{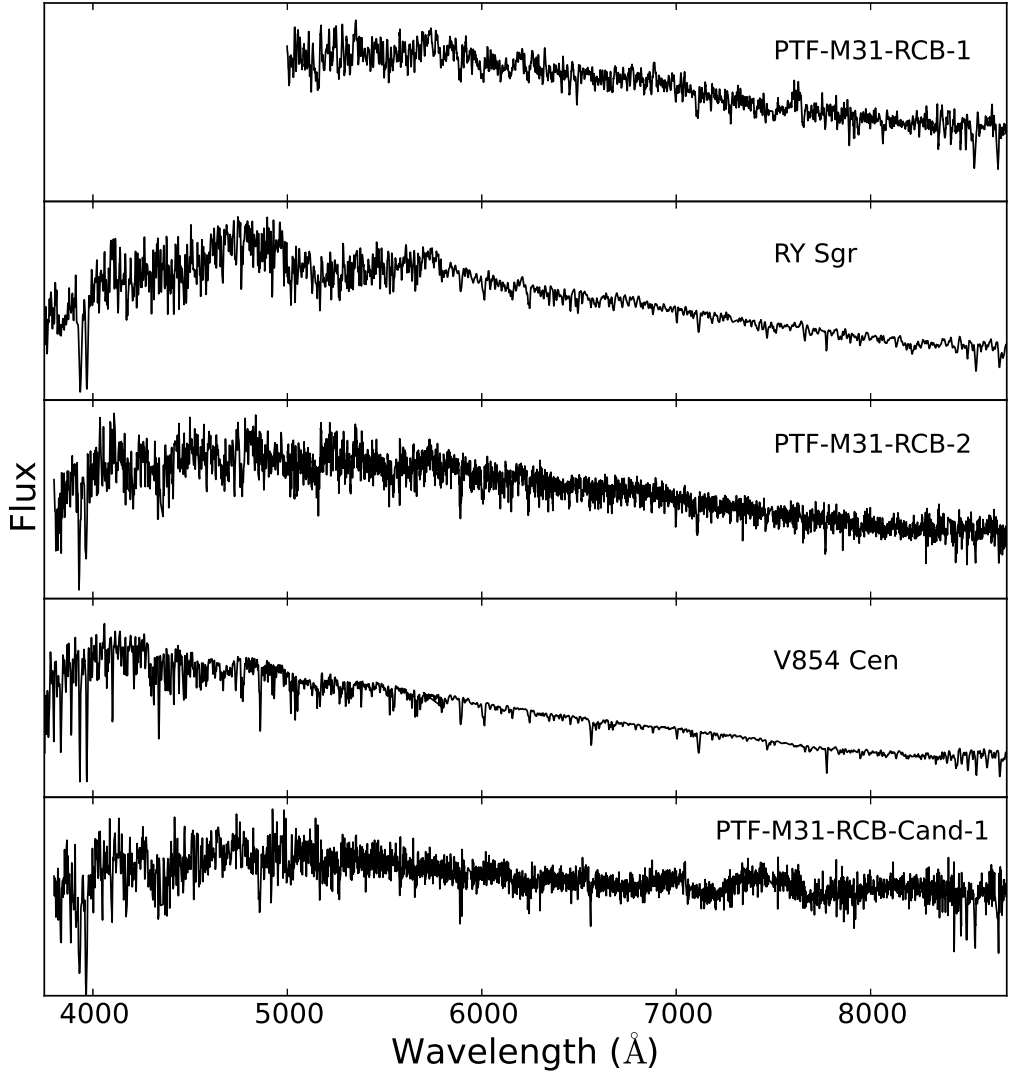}
\caption{Spectra of two newly confirmed RCBs and a RCB candidate in M31. The fluxes are arbitrarily normalized. 
For comparison, we also show spectra of two warm RCBs in the Milky Way, RY Sgr and V854 Cen (spectra from G. Clayton).
Expanded spectra in 5950-6750 \AA \ are shown in Figure 4.
\label{fig3}}
\end{figure}

\begin{figure}
\epsscale{1.0}
\plotone{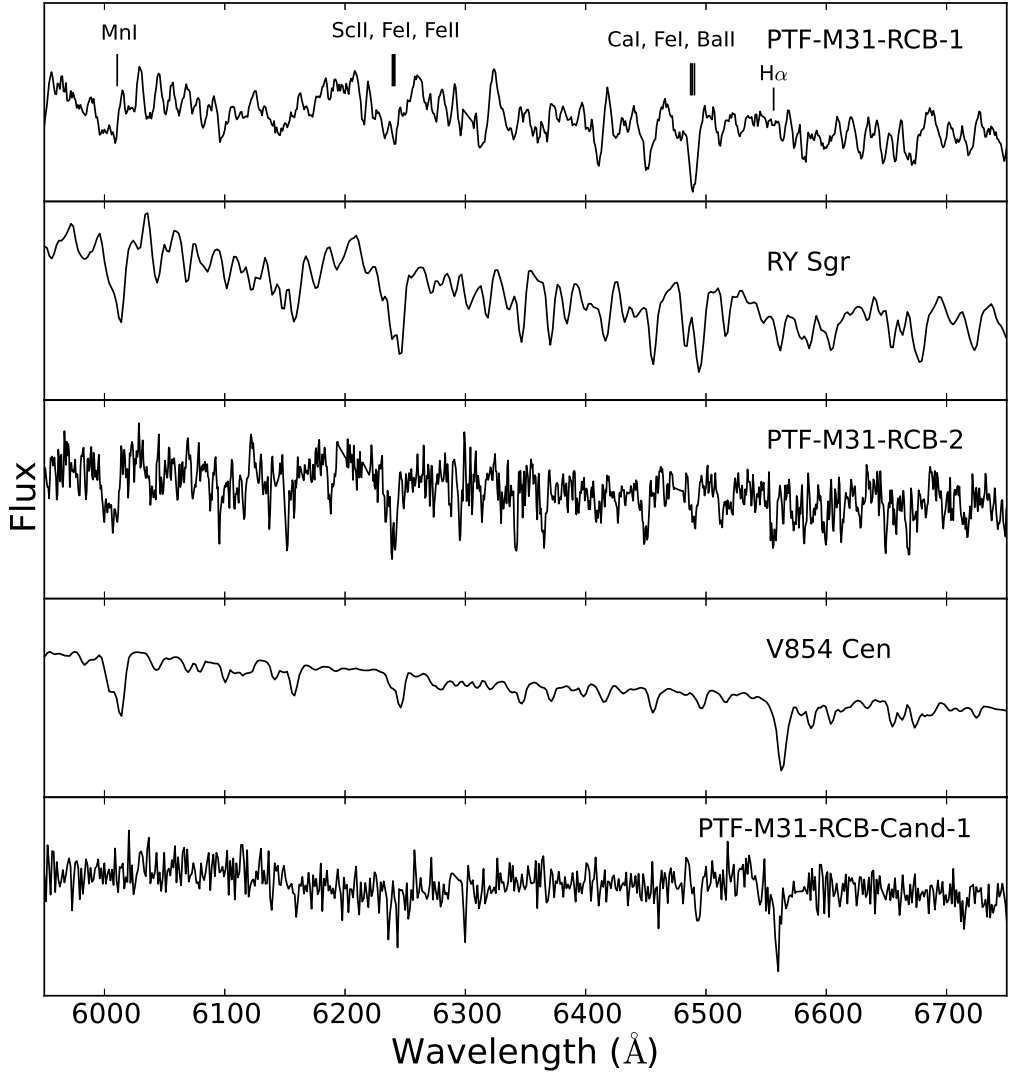}
\caption{A section of spectra of M31 RCBs in 5950-6750 \AA; Two Milky Way warm RCBs (RY Sgr and V854 Cen) are also plotted for comparison. 
Note the similarities between PTF-M31-RCB-1 (hereafter RCB-1, and so on) and RY Sgr, and RCB-2 and V854 Cen.
Some of the more prominent absorption lines (Danziger 1965) are marked in the upper-most panel, shifted by the radial velocity derived for RCB-1.
RCB-1 shows no sign of H$\rm \alpha$, while there are weak H$\rm \alpha$ absorption lines in RCB-2 and RCB-Cand-1.
\label{fig4}}
\end{figure}








\clearpage


\end{document}